\documentclass[12pt]{iopart}
\usepackage{graphicx}

\newcommand {\bx} {\mbox{\boldmath $x$}}

\newcommand{\calX}{{\cal X}}

\newcommand {\hx} {\hat{x}}
\newcommand {\hbx} {\hat{\mbox{\boldmath $x$}}}

\begin{document}

\title[A statistical--mechanical view on source coding]{A
statistical--mechanical view on source coding: physical compression and
data compression}

\author{Neri Merhav}

\address{Department of Electrical Engineering, Technion, Haifa 32000,
Israel.\\ E--mail: merhav@ee.technion.ac.il}

\begin{abstract}
We draw a certain analogy between the 
classical information--theoretic problem of lossy data compression
(source coding) of memoryless information sources
and the statistical mechanical behavior of a certain model of a chain
of connected particles (e.g., a polymer) that is subjected to a contracting
force. The free energy difference pertaining to such a contraction turns out
to be proportional to the rate--distortion function in the analogous data
compression model, and the contracting force is proportional to the derivative
this function. Beyond the fact that this analogy may be interesting on its own right,
it may provide a physical perspective on the behavior of optimum schemes for lossy data
compression (and perhaps also, an information--theoretic perspective on
certain physical
system models). Moreover, it triggers the derivation 
of lossy compression performance for systems with memory, using analysis tools
and insights from statistical mechanics.
\end{abstract}

\maketitle

\section{Introduction}

Relationships between information theory and
statistical physics have been widely recognized in the last few decades,
from a wide spectrum of aspects. These include conceptual aspects, of
parallelisms and analogies
between theoretical principles in
the two disciplines, as well as technical aspects, of
mapping between mathematical formalisms in both fields and borrowing
analysis techniques from one field to the other. One example
of such a mapping, is between the paradigm of random codes for channel coding
and certain models of magnetic materials,
most notably, Ising models and spin glass
models (see, e.g.,
\cite{Baierlein71},\cite{Merhav10},\cite{MM09},\cite{Nishimori01}, and many
references therein).
Today, it is quite widely believed
that research in the intersection between information theory and statistical
physics may have the potential of fertilizing both disciplines.

This paper is more related to the former aspect mentioned above, namely,
the relationships between the two areas in the conceptual level.
However, it has also ingredients from the second aspect.
In particular, let us consider two
questions in the two fields, which at first glance, may seem completely
unrelated, but will nevertheless turn out later to be very related. These are
special cases of more general questions that we study later in this paper.

\begin{figure}[ht]
\hspace*{1cm}\input{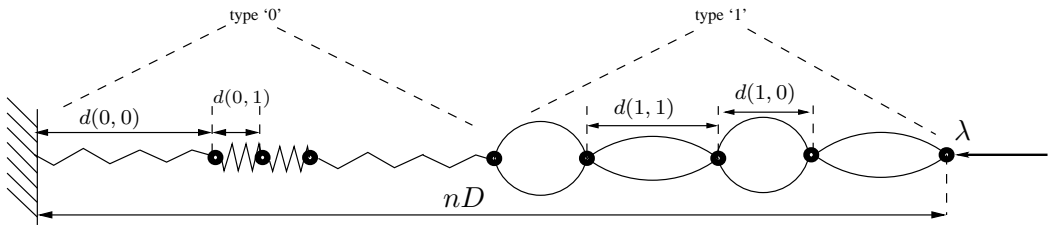}
\caption{A chain with various types of elements and various lengths.}
\label{chain}
\end{figure}

The first is a simple question in statistical mechanics, and it is about
a certain extension of a model described in \cite[page 134, Problem 13]{Kubo61}:
Consider a one--dimensional chain of $n$ connected elements (e.g., monomers or
whatever basic units that form a
polymer chain), arranged along a straight line (see Fig.\ \ref{chain}), 
and residing in thermal equilibrium at
fixed temperature $T_0$. The are two types of elements, which will be referred
to as type `0' and type `1'.
The number of elements of each type $x$ (with $x$ being either `0' or `1') is given by $n(x)=nP(x)$,
where $P(0)+P(1)=1$ (and so, $n(0)+n(1)=n$). 
Each element of each type may be in one of two different states, labeled by $\hx$, where
$\hx$ also takes on the values `0' and `1'.
The length and the internal energy of an element of type $x$ at state $\hx$
are given by  $d(x,\hx)$ and
$\epsilon(\hx)$ (independently of $x$), respectively. A contracting force
$\lambda < 0$ is applied to
one edge of the chain while the other edge is fixed. What is the minimum
amount of mechanical work $W$ that must be carried out by this force, along an isothermal
process at temperature $T_0$, in order 
to shrink the chain from its original
length $nD_0$ (when no force was applied) into a shorter length,
$nD$, where $D < 
D_0$ is a given constant? 

The second question is in information theory. In particular, it is
the classical problem of
lossy source coding, and some of the notation here will deliberately be chosen
to be the same as
before: An information source emits a string of $n$ independent
symbols, $x_1,x_2,\ldots,x_n$, where each $x_i$ may either be `0' or
`1', with probabilities $P(0)$ and $P(1)$, respectively. 
A lossy source encoder maps the source string, 
$(x_1,\ldots,x_n)$, into a shorter (compressed) representation
of average length $nR$, where $R$ is the 
coding rate (compression ratio), and the compatible decoder maps
this compressed representation into a reproduction string,
$\hbx=(\hx_1,\ldots,\hx_n)$, where each $\hx_i$ is again, either `0' or `1'.
The fidelity of the reproduction is measured in terms of a certain distortion
(or distance) function, $d(\bx,\hbx)=\sum_{i=1}^n d(x_i,\hx_i)$, which should
be as small as possible, so that $\hbx$ would be as `close' as possible to
$\bx$.\footnote{For example, in lossless compression, $\hbx$ is required to be
strictly identical to $\hbx$, in which case $d(\bx,\hbx)=0$. However, in some
applications, one might be willing to trade off between compression and fidelity,
i.e., slightly increase the distortion at the benefit of reducing the
compression ratio $R$.}
In the limit of large $n$, what is the minimum coding rate $R=R(D)$ 
for which there exists an encoder and decoder such that the average distortion,
$\left<d(\bx,\hbx)\right>$, would not exceed $nD$?

It turns out, as we shall see in the sequel,
that the two questions have intimately related answers. In particular,
the minimum amount of work $W$, in the first question, is related
to $R(D)$ (a.k.a.\ the {\it rate--distortion function}), of the second
question, according to
\begin{equation}
W=nkT_0R(D),
\end{equation}
provided that the Hamiltonian, $\epsilon(\hx)$, 
in the former problem, is given by
\begin{equation}
\epsilon(\hx)=-kT_0\ln Q(\hx),
\end{equation}
where $k$ is Boltzmann's constant, and $Q(\hx)$ is the 
relative frequency (or the empirical probability) of the symbol $\hx\in\{0,1\}$ in the
reproduction sequence $\hbx$, pertaining to an optimum lossy encoder--decoder
with average per--symbol distortion $D$ (for large $n$).
\begin{figure}[ht]
\hspace*{1cm}\input{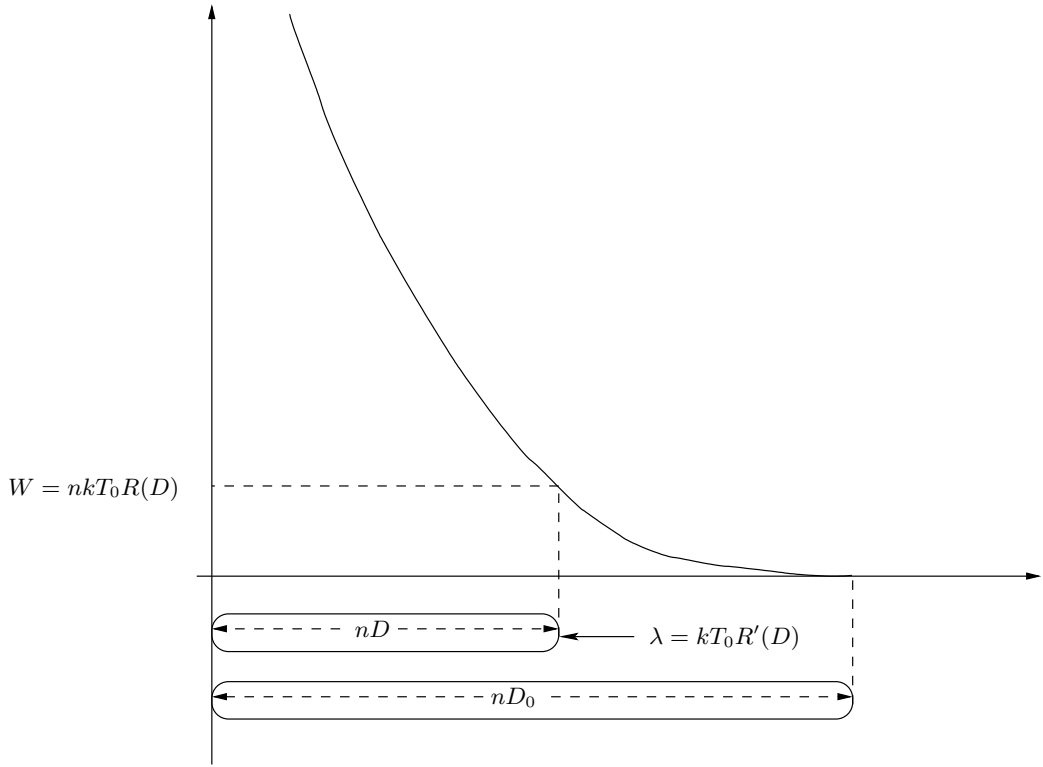}
\caption{Emulation of the rate--distortion function $R(D)$ by a physical system.}
\label{rd}
\end{figure}
Moreover, the minimum amount of work $W$, which is simply the free energy difference
between the final equilibrium state and the initial state of the chain, is achieved by a
reversible process, where the 
compressing force $\lambda$ grows very slowly from zero, at
the beginning of the process, up to a final level of 
\begin{equation}
\lambda = kT_0 R'(D),
\end{equation}
where $R'(D)$ is the derivative of $R(D)$ (see Fig.\ \ref{rd}).
Thus, physical compression is strongly related to data compression, and the
fundamental physical limit on the minimum required work is intimately
connected to the fundamental information--theoretic limit of the minimum
required coding rate. 

This link between the the physical model and the lossy source coding problem
is obtained from a large deviations perspective. 
The exact details will be seen later on, but in a nutshell, the idea is this:
On the one hand, it is
possible to represent $R(D)$ as the large deviations
rate function of a certain rare event, but on the other hand, this large
deviations rate function, involves the use of the Legendre transform, which
is a pivotal concept in thermodynamics and statistical mechanics. Moreover,
since this Legendre transform is applied to the (logarithm of the) moment generating
function (of the distortion variable), which in turn, has the form a partition
function, this paves the way to the above described analogy. The Legendre
transform is associated with the optimization across a certain parameter,
which can be interpreted as either inverse temperature (as was done, for
example, in
\cite{McAllester},\cite{Merhav08},\cite{Rose94},\cite{Shinzato}) or as a
(generalized) force, as proposed here. The interpretation of this parameter
as force is somewhat more solid, for reasons that will become apparent
later.

One application of this analogy, between the two models, is a 
parametric representation of the rate--distortion function $R(D)$ as
an integral of the minimum mean square error (MMSE) in a certain Bayesian estimation
problem, which is obtained in analogy to a certain variant of 
the fluctuation--dissipation theorem. This representation opens the door
for derivation of upper and lower bounds on the rate--distortion function via bounds on the MMSE,
as was demonstrated in a companion paper \cite{Merhav10b}. 

Another possible
application is demonstrated in the present
paper: When the setup is extended to allow information sources with memory
(non i.i.d.\ processes), then the analogous physical model consists of
interactions between the various particles. When these interactions are
sufficiently strong (and with high enough dimension), then the system exhibits
phase transitions. In the information--theoretic domain, these phase
transitions mean irregularities and threshold effects in the behavior of the
relevant information--theoretic function, in this case, the rate--distortion
function. Thus, analysis tools and physical insights are `imported' from
statistical mechanics to information theory.
A particular model example for this is worked out in Section 4.

The outline of the paper is as follows. In Section 2, we provide some
relevant background in information theory, which may safely be skipped 
by readers that possess this background. In Section 3, we establish the
analogy between lossy source coding and the above described physical model,
and discuss it in detail. In Section 4, we demonstrate the analysis for a
system with memory, as explained in the previous paragraph.
Finally, in Section 5 we summarize and conclude.

\section{Information--theoretic background}
\label{bkgd}

\subsection{General overview}

One of the most elementary roles of Information Theory is to provide fundamental
performance limits pertaining to certain tasks of information processing,
such as data compression, error--correction coding, encryption, data hiding, prediction,
and detection/estimation
of signals and/or parameters from noisy observations, just to name a few (see
e.g., \cite{CT06}). 

In this paper, our focus is on the first item mentioned --
data compression, a.k.a.\ {\it source coding}, 
where the mission is to convert a piece of information 
(say, a long file), henceforth referred to as the {\it source data}, into a shorter
(normally, binary) representation, which enables either perfect recovery of
the original information, as in the case of {\it lossless compression}, or non--perfect
recovery, where the level of reconstruction 
errors (or distortion) should remain within pre-specified
limits, which is the case of {\it lossy data compression}. 

Lossless compression is
possible whenever the statistical characterization of the
source data inherently exhibits some level of {\it redundancy} that
can be exploited by the compression scheme, for example, a binary file,
where the relative frequency of 1's is much larger than that of the 0's,
or when there is a strong statistical dependence between consecutive bits.
These types of redundancy exist, more
often than not, in real--life situations. If some level of errors and
distortion are allowed, as in the lossy case, then compression can be made
even more aggressive. 
The choice between lossless and lossy data compression depends on the
application and the type of data to be compressed. For example, when it comes
to sensitive information, like bank account information, or a piece of important text, 
then one may not tolerate any reconstruction errors at all. On the other hand, images and
audio/video files, may suffer some degree of harmless reconstruction errors (which may
be unnoticeable to the human eye or ear, if designed cleverly) and thus allow
stronger compression, which would be very welcome, since images and video
files are typically enormously large. The {\it compression ratio}, or the {\it
coding rate}, denoted $R$, is
defined as the (average) ratio between the length of the compressed file (in
bits) and the length of the original file. 

The basic role of Information Theory, in the context of lossless/lossy
source coding, is to characterize the fundamental limits of compression: For a
given statistical characterization of the source data, normally modeled by a certain
random process, what is the minimum achievable compression ratio $R$ as a function
of the allowed average distortion, denoted $D$, which is defined with respect 
to some distortion function that measures
the degree of proximity between the source data and the recovered data. 
The characterization of this minimum achievable $R$ for a given $D$,
denoted as a function $R(D)$, is called the {\it rate--distortion function} of
the source with respect to the prescribed distortion function.
For the lossless case, of course, $D=0$.
Another important question is how, in principle, one may achieve (or at least
approach) this fundamental limit of optimum performance, $R(D)$? In this context, there is a big
gap between lossy compression and lossless compression. While for the lossless case,
there are many practical algorithms (most notably, adaptive Huffman codes, Lempel--Ziv
codes, arithmetic codes, and more), in the lossy case, there is unfortunately,
no constructive practical scheme whose performance comes close to $R(D)$.

\subsection{The rate--distortion function}
The simplest non--trivial model of an information source is that of an i.i.d.\
process, a.k.a.\ a {\it discrete memoryless source} (DMS), where the source
symbols, $x_1,x_2,\ldots,x_n$, take on values in a common finite set (alphabet)
$\calX$, they are
statistically independent, and they are all drawn
from the same probability mass function, denoted by $P=\{P(x),~x\in\calX\}$. 
The source string $\bx=(x_1,\ldots,x_n)$ is compressed into a binary
representation\footnote{
It should be noted that in the case of
variable--length coding, where $\ell=\ell(\bx)$ depends on $\bx$, the code
should be designed such that the running bit-stream (formed by 
concatenating compressed strings corresponding
to successive $n$--blocks from the source) could be uniquely parsed in the correct manner
and then decoded.
To this end, the lengths $\{\ell(\bx)\}$ must be collectively large enough so as to satisfy the
Kraft inequality. The details can be found, for example, in \cite{CT06}.}
of length $\ell=\ell(\bx)$ (which may or may not depend on $\bx$), whose
average is $\left<\ell(\bx)\right>$, and the compression ratio is
$R=\left<\ell(\bx)\right>/n$. 
In the decoding (or decompression) process, the
compressed representation is mapped into a reproduction string
$\hbx=(\hx_1,\hx_2,\ldots,\hx_n)$, where each $\hx_i$, $i=1,2,\ldots,n$, takes
on values in the {\it reproduction alphabet} $\hat{\calX}$ (which is typically
either equal to $\calX$ or to a subset of $\calX$, but this is not necessary). 
The fidelity of the reconstruction string $\hbx$ relative to the original source
string $\hx$ is
measured by a certain distortion function 
$d_n(\bx,\hbx)$, where the function $d_n$ is defined
additively as $d_n(\bx,\hbx)=\sum_{i=1}^n d(x_i,\hx_i)$, $d(\cdot, \cdot)$
being a function from $\calX\times\hat{\calX}$ to the non--negative reals.
The average distortion per symbol is $D=\left<d_n(\bx,\hbx)\right>/n$.

As said, $R(D)$ is defined (in general) as the infimum of all rates $R$ for which there exist a
sufficiently large $n$ and an encoder--decoder pair for $n$--blocks,
such that the average distortion per symbol would not exceed $D$.
In the case of a DMS $P$, an elementary coding theorem of Information Theory
asserts that $R(D)$ admits the following formula
\begin{equation}
\label{rdc}
R(D)=\min I(x;\hat{x}),
\end{equation}
where $x$ is a random variable that represents a single source symbol (i.e., it is
governed by $P$),
$I(x;\hx)$ is the mutual information between $x$ and $\hx$, i.e.,
\begin{equation}
I(x;\hx)=\left<\log\frac{W(\hx|x)}{Q(\hx)}\right>\equiv
\sum_{x\in\calX}\sum_{\hx\in\hat{\calX}}P(x)P(\hx|x)\log\frac{W(\hx|x)}{Q(\hx)},
\end{equation}
$Q(\hx)=\sum_{x\in\calX}P(x)W(\hx|x)$ being the marginal distribution
of $\hx$, which is associated with a given conditional distribution
$\{W(\hx|x)\}$,
and the minimum is over all these conditional probability
distributions for which 
\begin{equation}
\left<d(x,\hx)\right>\equiv
\sum_{x\in\calX}\sum_{\hx\in\hat{\calX}}P(x)W(\hx|x)d(x,\hx) \le D.
\end{equation}
For $D=0$, $\hx$ must be equal to $x$ with probability one (unless
$d(x,\hx)=0$ also for some $\hx\ne x$), and then
\begin{equation}
R(0)=I(x;x)=-\left<\log P(x)\right>\equiv H, 
\end{equation}
the Shannon entropy of $x$, as expected.
As mentioned earlier, there are concrete compression algorithms 
that come close to $H$ for large $n$.
For $D > 0$, however, the proof of achievability of $R(D)$ is
non--constructive. 

\subsection{Random coding}

The idea for proving the existence of a sequence of codes
(indexed by $n$) whose performance approach $R(D)$ as $n\to\infty$, is based on the notion of
{\it random coding}: If we can define, for each $n$, an ensemble of codes of (fixed) rate
$R$, for which the average per--symbol distortion (across both the randomness of $\hx$ and
the randomness of the code) is asymptotically less than or equal to $D$, 
then there must exist at least one
sequence of codes in that ensemble, with this property. The idea of random
coding is useful because if the ensemble of codes is chosen wisely, the average ensemble
performance is surprisingly easy to derive (in contrast to the performance of
a specific code) and proven to meet $R(D)$ in the limit of large $n$.

For a given $n$, consider the following ensemble of codes: Let $W^*$ denote the
conditional probability matrix that achieves $R(D)$ and let $Q^*$ denote the
corresponding marginal distribution of $\hx$. Consider now a random selection
of $M=e^{nR}$ reproduction strings, $\hbx_1,\hbx_2,\ldots,\hbx_M$, each of
length $n$, where each $\hbx_i=(\hx_{i,1},\hx_{i,2},\ldots,\hx_{i,n})$,
$i=1,2,\ldots,M$,
is drawn independently (of all other reproduction
strings), according to
\begin{equation}
Q^*(\hbx_i)=Q^*(\hx_{i,1})Q^*(\hx_{i,2})\cdot\cdot\cdot Q^*(\hx_{i,n}).
\end{equation}
This randomly chosen code is generated only once and then revealed to the
decoder. Upon observing an incoming source string $\bx$, the encoder seeks the
first reproduction string $\hbx_i$ that achieves $d_n(\bx,\hbx_i)\le nD$, and
then transmits its index $i$ using $\log_2 M=nR\log_2e$ bits, or equivalently,
$\ln M=nR$ {\it nats}.\footnote{While $\log_2M$ has the obvious interpretation of the
number of bits needed to specify a number between $1$ and $M$, the natural
base logarithm is often mathematically more convenient to work with. The
quantity $\ln M$ can also
be thought of as the description length, but in different units, called nats, rather
than bits, where the conversion is according to $1$ nat $=\log_2 e$ bits.} If
no such codeword exists, which is referred to as the
event of {\it encoding failure}, the encoder sends an arbitrary sequence of
$nR$ nats, say, the all--zero sequence.
The decoder receives the index $i$ and simply outputs the corresponding reproduction
string $\hbx_i$. 

Obviously, the per--symbol distortion would be less than $D$
whenever the encoder does not fail, and so, the main point of the proof is to
show that the probability of failure (across the 
randomness of $\bx$ and the ensemble of codes) 
is vanishingly small for large $n$,
provided that $R$ is slightly larger than (but 
can be arbitrarily close to) $R(D)$, i.e., $R=R(D)+\epsilon$ for
an arbitrarily small $\epsilon > 0$. The idea is that for any source string
that is {\it typical} to $P$ (i.e., the empirical relative frequency of
each symbol in $\bx$ is close to its probability), one can show (see, e.g.,
\cite{CT06}) that the
probability that a single, randomly selected reproduction string $\hbx$
would satisfy $d_n(\bx,\hbx)\le nD$, decays exponentially as $\exp[-nR(D)]$.
Thus, the above described random selection of the entire codebook together
with the encoding operation, are
equivalent to conducting $M$ independent trials in the quest for having at
least one $i$ for which 
$d_n(\bx,\hbx_i)\le nD$, $i=1,2,\ldots,M$. If $M = e^{n[R(D)+\epsilon]}$, the
number of trials is much larger (by a factor of $e^{n\epsilon}$) 
than the reciprocal of the probability of
a single `success', $\exp[-nR(D)]$, and so, the probability of obtaining at
least one such success (which is case where the encoder succeeds) tends to unity as
$n\to\infty$. We took the liberty of assuming that source string is typical to
$P$ because the probability of seeing a non--typical
string is vanishingly small.

\subsection{The Large deviations perspective}

From the foregoing discussion, 
we see that
$R(D)$ has the additional interpretation of the exponential rate
of the probability of the event
$d_n(\bx,\hbx)\le nD$, where $\bx$ is a given string typical to $P$ and $\hbx$
is randomly drawn i.i.d.\ under $Q^*$. Consider the following chain of
equalities and inequalities for bounding the probability of this event from above.
Letting $s$ be a parameter taking an arbitrary non--positive value, we have:
\begin{eqnarray}
\mbox{Pr}\left\{d_n(\bx,\hbx)\le nD\right\}&=&
\mbox{Pr}\left\{\sum_{i=1}^n d(x_i,\hx_i)\le nD\right\}\nonumber\\
&\le&\left<\exp\left\{s\left[\sum_{i=1}^n
d(x_i,\hx_i)-nD\right]\right\}\right>\nonumber\\
&=&e^{-nsD}\left<\prod_{i=1}^n e^{sd(x_i,\hx_i)}\right>\nonumber\\
&=&e^{-nsD}\prod_{i=1}^n\left<e^{sd(x_i,\hx_i)}\right>\nonumber\\
&=&e^{-nsD}\prod_{x\in\calX}\prod_{i:~x_i=x}\left<e^{sd(x,\hx_i)}\right>\nonumber\\
&=&e^{-nsD}\prod_{x\in\calX}\left[\left<e^{sd(x,\hx)}\right>\right]^{nP(x)}\nonumber\\
&=&e^{-nI(D,s)}
\end{eqnarray}
where $I(D,s)$ is defined as
\begin{equation}
I(D,s)=\exp\left\{-n\left[sD-\sum_{x\in\calX}P(x)\ln\left(
\sum_{\hx\in\hat{\calX}}Q^*(\hx)e^{sd(x,\hx)}\right)\right]\right\}.
\end{equation}
The tightest upper bound is obtained by minimizing it over the range $s\le 0$,
which is equivalent to maximizing $I(D,s)$ in that range. I.e., the tightest
upper bound of this form is $e^{-nI(D)}$, where $I(D)=\sup_{s\le 0}I(D,s)$
(the Chernoff bound).
While this is merely an upper bound, the methods of large deviations theory
(see, e.g., \cite{DZ93}) can readily be used to establish the fact that the
bound $e^{-nI(D)}$
is tight in the exponential sense, namely, it is the correct asymptotic exponential
decay rate of $\mbox{Pr}\{d_n(\bx,\hbx)\le nD\}$. Accordingly,
$I(D)$ is called the {\it large deviations rate function} of this event. Combining this with the
foregoing discussion, it follows that $R(D)=I(D)$, which means that an
alternative expression of $R(D)$ is given by
\begin{equation}
\label{ldrdo}
R(D)=\sup_{s\le
0}\left[sD-\sum_{x\in\calX}P(x)\ln\left(
\sum_{\hx\in\hat{\calX}}Q^*(\hx)e^{sd(x,\hx)}\right)\right].
\end{equation}
Interestingly, the same expression was obtained in \cite[Corollary
4.2.3]{Gray90} using completely different considerations (see also
\cite{Rose94}). In this paper,
however, we will also concern ourselves, more generally, with the rate--distortion
function, $R_Q(D)$, pertaining to a given reproduction distribution $Q$, which may
not necessarily be the optimum one, $Q^*$. This function is defined similarly as in
eq.\ (\ref{rdc}), but with the additional constraint that the marginal 
distribution that represents the reproduction would agree with the given $Q$,
i.e., $\sum_{x}P(x)W(\hx|x)=Q(\hx)$.
By using the same large deviations
arguments as above, but for an arbitrary random coding distribution $Q$, one 
readily observes that $R_Q(D)$ is of the same form as in eq.\ (\ref{ldrdo}),
except that $Q^*$ is replaced by the given $Q$ (see also \cite{Merhav10b}).
This expression will now be used as a bridge to the realm of equilibrium
statistical mechanics.

\section{Statistical mechanics of source coding}

Consider the parametric representation of the rate--distortion function
$R_Q(D)$, with respect to a given reproduction distribution $Q$:
\begin{equation}
\label{ldrd}
R_Q(D)=\sup_{s\le
0}\left[sD-\sum_{x\in\calX}P(x)\ln\left(
\sum_{\hx\in\hat{\calX}}Q(\hx)e^{sd(x,\hx)}\right)\right].
\end{equation}
The expression in the inner brackets,
\begin{equation}
Z_x(s)\equiv\sum_{\hx\in\hat{\calX}}Q(\hx)e^{sd(x,\hx)},
\end{equation}
can be thought of as the partition function 
of a single particle of ``type'' $x$, which is defined as follows.
Assuming a certain fixed temperature $T=T_0$, consider the Hamiltonian
\begin{equation}
\epsilon(\hx)=-kT_0\ln Q(\hx).
\end{equation}
Imagine now that this particle may be in various 
states, indexed by $\hx\in\hat{\calX}$. When a particle of type $x$ lies in state
$\hx$ its internal energy is $\epsilon(\hx)$, as defined above, and its length
is $d(x,\hx)$. Next assume that instead of working with the 
parameter $s$, we rescale and redefine 
the free parameter as $\lambda$, where $s=\lambda/(kT_0)$. Then,
$\lambda$ has the physical meaning of a force that is conjugate to
the length. This force is stretching for $\lambda > 0$
and contracting for $\lambda < 0$. With a slight abuse of notation, the
Gibbs partition function \cite[Section 4.8]{Kardar07} 
pertaining to a single particle of type $x$ is then given by
\begin{equation}
Z_x(\lambda)=\sum_{\hx\in\hat{\calX}}\exp\left\{-\frac{1}{kT_0}[\epsilon(\hx)-\lambda
d(x,\hx)]\right\},
\end{equation}
and accordingly,
\begin{equation}
G_x(\lambda)=-kT_0\ln Z_x(\lambda)
\end{equation}
is the Gibbs free energy per particle of type $x$. Thus,
\begin{equation}
G(\lambda)=\sum_{x\in\calX}P(x)G_x(\lambda)
\end{equation}
is the average per--particle Gibbs free energy (or the Gibbs free energy density) 
pertaining to a system with a total of
$n$ non--interacting particles, from $|\calX|$ different types, where the number
of particles of type $x$ is $nP(x)$, $x\in\calX$.
The Helmholtz free energy per particle is then given by the Legendre transform
\begin{equation}
\label{legendre}
F(D)=\sup_{\lambda}[G(\lambda)+\lambda D].
\end{equation}
However, for $D < D_0\equiv \sum_{x\in\calX}\sum_{\hx\in\hat{\calX}}P(x)Q(\hx)d(x,\hx)$
(which is the interesting range, where $R_Q(D) > 0$), the maximizing $\lambda$
is always non--positive, and so,
\begin{equation}
F(D)=\sup_{\lambda \le 0}[G(\lambda)+\lambda D].
\end{equation}
Invoking now eq.\ (\ref{ldrd}), we readily identify that
\begin{equation}
F(D)=kT_0R_Q(D),
\end{equation}
which supports the analogy between the lossy data compression problem and
the behavior of the statistical--mechanical model of the kind
described in the third paragraph of the Introduction: According
to this model, the physical system under discussion is a long chain
with a total of $n$ elements, which is composed of $|\calX|$ different types of
shorter chains (indexed by $x$), where the number of elements in the short
chain of type $x$
is $nP(x)$, and where each element of each chain can be in various states,
indexed by $\hx$. In each state $\hx$, the internal 
energy and the length of each element are $\epsilon (\hx)$ and $d(x,\hx)$, 
as described above. The total length
of the chain, when no force is applied, is therefore
$\sum_{i=1}^n\left<d(x_i,\hx_i)\right>|_{\lambda=0}=nD_0$. Upon applying a contracting
force $\lambda < 0$, 
states of shorter length become
more probable, and the chain shrinks 
to the length of $nD$, where $D$ is related to $\lambda$ according to
the Legendre relation\footnote{
Since $G(\lambda)$ is concave and $F(D)$ is convex, 
the inverse Legendre transform holds as well, and so, there is one--to--one
correspondence between $\lambda$ and $D$.}
(\ref{legendre}) between $F(D)$ and $G(\lambda)$, which is given by
\begin{equation}
\label{fl}
\lambda = F'(D)=kT_0R_Q'(D),
\end{equation}
where $F'(D)$ and $R_Q'(D)$ are, respectively, the derivatives of $F(D)$ and 
$R_Q(D)$ relative to $D$. The inverse relation is, of course,
\begin{equation}
D=-G'(\lambda),
\end{equation}
where $G'(\lambda)$ is the derivative of $G(\lambda)$. Since $R_Q(D)$ is
proportional to the free energy, where the system is held in equilibrium at
length $nD$, it also means the minimum amount of work required in order
to shrink the system from length $nD_0$ to length $nD$, and this minimum is
obtained by a reversible process of slow increase in $\lambda$, starting from
zero and ending at the final value given by eq.\ (\ref{fl}).\\

\noindent
{\it Discussion}\\
This analogy between the lossy source coding problem and the
statistical--mechanical model of a chain, may suggest that physical insights
may shed light on lossy source coding and vice versa. We learn, for
example, that the contribution of each source symbol $x$ to the
distortion, $\sum_{i:~x_i=x}d(x_i,\hx_i)$, is analogous to the 
length contributed by the chain of type $x$ when the contracting force $\lambda$ is
applied. We have also learned that the local slope of $R_Q(D)$ is
proportional to a force, which must increase as the chain is contracted more
and more aggressively, and
near $D=0$, it normally tends to infinity, as $R_Q'(0)=-\infty$ in most cases.
This slope parameter also plays a pivotal role in theory and practice of lossy
source coding: On the theoretical side, it gives rise to a variety of
parametric representations of the rate--distortion 
function \cite{Berger71},\cite{Gray90}, some of
which support the derivation of important, non--trivial bounds. On the more
practical side, often data compression schemes are designed by optimizing
an objective function with the structure of
$$\mbox{rate}+\lambda\cdot\mbox{distortion},$$
thus $\lambda$ plays the role of a Lagrange multiplier.
This Lagrange multiplier is now understood to act like a physical force,
which can be `tuned' to the desired trade--off between rate and distortion.
As yet another example, the convexity of 
the rate--distortion function can be understood from a
physical point of view, as the Helmholtz free energy is also convex, a fact
which has a physical explanation (related to the fluctuation--dissipation
theorem), in addition to the mathematical one.

At this point, two technical comments are in order:
\begin{enumerate}
\item We emphasized the fact that the reproduction distribution $Q$ is fixed.
For a given target value of $D$, one may, of course, have the freedom to
select the optimum distribution $Q^*$ that minimizes $R_Q(D)$, which would
yield the rate--distortion function, $R(D)$, and so, in principle, all the
foregoing discussion applies to $R(D)$ as well. Some caution, however, must
be exercised here, because in general, the optimum $Q$ may depend on $D$ (or
equivalently, on $s$ or $\lambda$), which means, that in the
analogous physical model, the internal energy $\epsilon(\hx)$ depends on the
force $\lambda$ (in addition to the linear dependence of the term
$\lambda d(x,\hx)$). This kind of dependence does not support the above described
analogy in a natural manner. This is the reason that we have defined the
rate---distortion problem for a fixed $Q$, as it avoids this problem.
Thus, even if we pick the optimum
$Q^*$ for a given target distortion level $D$, then this $Q^*$ must be
kept unaltered throughout the entire process of increasing $\lambda$ from zero
to its final value, given by (\ref{fl}), although $Q^*$ may be sub--optimum for
all intermediate distortion values that are met along the way from $D_0$ to $D$.
\item An alternative interpretation of the parameter $s$, in the
partition function $Z_x(s)$, could be the (negative) inverse
temperature, as was suggested in \cite{Rose94} (see also \cite{Merhav08}). 
In this case, $d(x,\hx)$ would be the internal energy of an element of type
$x$ at state $\hx$ and $Q(\hx)$, which does not include a power of $s$, 
could be understood as being proportional to the degeneracy (in some
coarse--graining process). In this case, the distortion would have the meaning
of internal energy, and since no mechanical work is involved, this would also
be the heat absorbed in the system, 
whereas $R_Q(D)$ would be related to the entropy of the system.
The Legendre transform, in this case, is the one pertaining to the passage
between the microcanonical ensemble and the canonical one.
The advantage of the interpretation of $s$ (or $\lambda$) as force, as
proposed here, is that it lends itself naturally to
a more general case, where there
is more than one fidelity criterion. For example, suppose there are two fidelity criteria,
with distortion functions $d$ and $d'$. Here, there would be two conjugate
forces, $\lambda$ and $\lambda'$, respectively 
(for example, a mechanical force and a magnetic force), and the physical analogy
carries over. On the other hand, this would not work naturally with the
temperature interpretation approach since there is only one temperature parameter in
physics.
\end{enumerate}

We end this section by providing a representation of $R_Q(D)$ and $D$ in an integral form,
which follows as a simple consequence of its representation as the Legendre
transform of $\ln Z_x(s)$, as in eq.\ (\ref{ldrd}).
Since the maximization problem in (\ref{ldrd}) is a convex problem ($\ln Z_x(s)$ is convex in
$s$), the minimizing $s$ for a given $D$ is obtained by taking the derivative
of the r.h.s., which leads to
\begin{eqnarray}
D&=&\sum_{x\in\calX}P(x)\cdot\frac{\partial\ln Z_x(s)}{\partial s}\nonumber\\
&=&\sum_{x\in\calX}P(x)\cdot\frac{\sum_{\hx\in\hat{\calX}}Q(\hx)d(x,\hx)e^{sd(x,\hx)}}
{\sum_{\hx\in\hat{\calX}}Q(\hx)e^{sd(x,\hx)}}.
\end{eqnarray}
This equation yields the distortion level $D$ for a given value of the
minimizing $s$ in eq.\ (\ref{ldrd}). Let us then denote
\begin{equation}
\label{ds}
D_s\equiv\sum_{x\in\calX}P(x)\cdot\frac{\sum_{\hx\in\hat{\calX}}Q(\hx)d(x,\hx)e^{sd(x,\hx)}}
{\sum_{\hx\in\hat{\calX}}Q(\hx)e^{sd(x,\hx)}},
\end{equation}
which means that
\begin{equation}
\label{rds}
R_Q(D_s)=sD_s-\sum_{x\in\calX}P(x)\ln Z_x(s).
\end{equation}
Taking the derivative of (\ref{ds}), we readily obtain
\begin{eqnarray}
\label{derds}
\frac{\mbox{d}D_s}{\mbox{d}s}&=&\sum_{x\in\calX}P(x)\frac{\partial}{\partial
s}\left[\frac{\sum_{\hx\in\hat{\calX}}Q(\hx)d(x,\hx)e^{sd(x,\hx)}}
{\sum_{\hx\in\hat{\calX}}Q(\hx)e^{sd(x,\hx)}}\right]\nonumber\\
&=&\sum_{x\in\calX}P(x)
\left[\frac{\sum_{\hx\in\hat{\calX}}Q(\hx)d^2(x,\hx)e^{sd(x,\hx)}}
{\sum_{\hx\in\hat{\calX}}Q(\hx)e^{sd(x,\hx)}}-\right.\nonumber\\
& &\left.\left(\frac{\sum_{\hx\in\hat{\calX}}Q(\hx)d(x,\hx)e^{sd(x,\hx)}}
{\sum_{\hx\in\hat{\calX}}Q(\hx)e^{sd(x,\hx)}}\right)^2\right]\nonumber\\
&=&\sum_{x\in\calX}P(x)\cdot\mbox{Var}_s\{d(x,\hx)|x\}\nonumber\\
&\equiv&\mbox{mmse}_s\{d(x,\hx)|x\},
\end{eqnarray}
where $\mbox{Var}_s\{d(x,\hx)|x\}$ is the variance of $d(x,\hx)$ w.r.t.\
the conditional probability distribution 
\begin{equation}
W_s(\hx|x)=\frac{Q(\hx)e^{sd(x,\hx)}}
{\sum_{\tilde{x}\in\hat{\calX}}Q(\tilde{x})e^{sd(x,\tilde{x})}}.
\end{equation}
The last line of eq.\ (\ref{derds}) means that the
expectation of $\mbox{Var}_s\{d(x,\hx)|x\}$ w.r.t.\ $P$ is exactly the
MMSE of estimating $d(x,\hx)$ based on the `observation' $x$ using the conditional mean
of $d(x,\hx)$ given $x$ as an
estimator. Differentiating both sides of eq.\ (\ref{rds}), we get
\begin{eqnarray}
\frac{\mbox{d}R_Q(D_s)}{\mbox{d}s}&=&s\cdot\frac{\mbox{d}D_s}{\mbox{d}s}+D_s-
\sum_{x\in\calX}P(x)\cdot\frac{\partial\ln Z_x(s)}{\partial s}\nonumber\\
&=&s\cdot\mbox{mmse}_s\{d(x,\hx)|x\}+D_s-D_s\nonumber\\
&=&s\cdot\mbox{mmse}_s\{d(x,\hx)|x\},
\end{eqnarray}
or, equivalently, 
\begin{equation}
R_Q(D_s)=\int_0^s s'\cdot \mbox{mmse}_{s'}\{d(x,\hx)|x\}\mbox{d}s',
\end{equation}
and
\begin{equation}
D_s=D_0+\int_0^s\mbox{mmse}_{s'}\{d(x,\hx)|x\}\mbox{d}s'.
\end{equation}
In \cite{Merhav10b}, this representation was studied
extensively and was found quite useful. 
In particular, simple bounds on the MMSE were shown
to yield non--trivial bounds on the rate--distortion function in some
cases where an exact closed form expression is unavailable. The physical
analogue of this representation is the fluctuation--dissipation theorem,
where the conditional variance, or equivalently the MMSE, plays the
role of the fluctuation, which describes the sensitivity, or the linear
response, of the length of the system to a small perturbation in the
contracting force. If $s$ is interpreted as the negative inverse temperature,
as was mentioned before, then the MMSE is related to the specific heat of the
system.

\section{Sources with memory and interacting particles}

The theoretical framework established in the previous section extends, in
principle, to information sources with memory 
(non i.i.d.\ sources), with a natural correspondence to a physical
system of interacting particles. While the rate--distortion function for a
general source with memory is unknown, the maximum rate achievable by 
random coding can still be derived in many cases of interest. 
Unlike the case of the memoryless source, where the best random coding
distribution is memoryless as well, when the source exhibits
memory, there is no apparent reason to believe that good random coding
distributions should remain memoryless either, but it is not known what
the form of the optimum random coding distribution is. For example, there is
no theorem that asserts that the optimum random coding distribution for
a Markov source is Markov too. One can, however examine various forms
of the random coding distributions and compare them. Intuitively, the stronger
is the memory of the source, the stronger should be the memory of the random
coding distribution. 

In this section, we demonstrate one family of random coding distributions, with a
very strong memory, which is inspired by the Curie--Weiss model of spin
arrays, that possesses long range interactions. Consider the random coding distribution
\begin{equation}
Q(\hbx)=\frac{\exp\left\{B\sum_{i=1}^n
\hx_i+\frac{J}{2n}\left(\sum_{i=1}^n\hx_i\right)^2\right\}}{Z_n(B,J)}
\end{equation}
where $\hat{\calX}=\{-1,+1\}$, $B$ and $J$ are parameters, and $Z_n(B,J)$ is
the appropriate normalization constant.
Using the identity,
\begin{equation}
\exp\left\{\frac{J}{2n}\left(\sum_{i=1}^n\hx_i\right)^2\right\}=
\sqrt{\frac{n}{2\pi J}}\int_{-\infty}^{+\infty}\mbox{d}\theta
\exp\left\{-\frac{n\theta^2}{2J}+\theta\sum_{i=1}^n\hx_i\right\},
\end{equation}
we can represent $Q$ as a mixture of i.i.d.\ distributions as follows:
\begin{equation}
Q(\hbx)=\int_{-\infty}^{+\infty}\mbox{d}\theta \pi_n(\theta)Q_{\theta}(\hbx)
\end{equation}
where $Q_\theta$ is the memoryless source
\begin{equation}
Q_{\theta}(\hbx)=\frac{\exp\{(B+\theta)\sum_{i=1}^n\hx_i\}}{[2\cosh(B+\theta)]^n}
\end{equation}
and the weighting function $\pi_n(\theta)$ is given by
\begin{equation}
\pi_n(\theta)=\frac{1}{Z_n(B,J)}\sqrt{\frac{n}{2\pi
J}}\exp\left\{-n\left[\frac{\theta^2}{2J}-\ln[2\cosh(B+\theta)]\right]\right\}.
\end{equation}
Next, we repeat the earlier derivation for each $Q_\theta$ individually:
\begin{eqnarray}
Q\left\{\sum_i d(x_i,\hx_i)\le nD\right\}&=&
\int_{-\infty}^{+\infty}\mbox{d}\theta
\pi_n(\theta)Q_{\theta}\left\{\sum_{i=1}^nd(x_i,\hx_i)\le
nD\right\}\nonumber\\
&\le&
\int_{-\infty}^{+\infty}\mbox{d}\theta
\pi_n(\theta)e^{-nR_{\theta}(D)},
\end{eqnarray}
where $R_\theta(D)$ is a short--hand notation for $R_{Q_\theta}(D)$, which is
well defined from the previous section since $Q_\theta$ is an i.i.d.\ distribution.
At this point, two observations are in order: First,
we observe that a separate large deviations analysis for each
i.i.d.\ component $Q_\theta$ is better than applying a similar analysis
directly to $Q$ itself,
without the decomposition, since it allows a different optimum choice of $s$
for each $\theta$, rather than one optimization of $s$ that compromises all values of
$\theta$. Moreover, since the upper bound is exponentially tight for
each $Q_\theta$, then the corresponding mixture of bounds is also
exponentially tight. The second observation is that 
since $Q_\theta$ is i.i.d., $R_{\theta}(D)$ depends on
the source $P$ only via the marginal distribution of a single symbol
$P(x)=\mbox{Pr}\{x_i=x\}$, which is assumed here to be independent of $i$.

A saddle--point analysis gives rise to the following
expression for $R_Q(D)$, the random--coding rate distortion function
pertaining to $Q$, which is the large deviations rate function:
\begin{equation}
R_Q(D)=\min_{\theta}\left\{\frac{\theta^2}{2J}-\ln[2\cosh(B+\theta)]+R_{\theta}(D)\right\}
+\phi(B,J)
\end{equation}
where
\begin{equation}
\phi(B,J)=\lim_{n\to\infty}\frac{\ln Z_n(B,J)}{n}.
\end{equation}
We next have a closer look at $R_{\theta}(D)$, assuming
$\calX=\hat{\calX}=\{-1,+1\}$,
and using the Hamming distortion
function, i.e.,
\begin{equation}
d(x,\hx)=\frac{1-x\cdot\hx}{2}=\left\{\begin{array}{ll}
0 & x=\hx\\
1 & x\ne \hx \end{array}\right.
\end{equation}
Since
\begin{eqnarray}
\sum_{\hx}Q_{\theta}(\hx)e^{sd(x,\hx)}&=&
\sum_{\hx}\frac{e^{(B+\theta)\hx}}{2\cosh(B+\theta)}\cdot
e^{s(1-x\hx)/2}\nonumber\\
&=&\frac{e^{s/2}\cosh(B+\theta-sx/2)}{\cosh(B+\theta)},
\end{eqnarray}
we readily obtain
\begin{eqnarray}
R_\theta(D)&=&\max_{s\le
0}\left[s\left(D-\frac{1}{2}\right)-\sum_xP(x)
\ln\cosh\left(B+\theta-\frac{sx}{2}\right)\right]\nonumber\\
& &+\ln\cosh(B+\theta).
\end{eqnarray}
On substituting this expression back into the expression of $R_Q(D)$, we
obtain the formula
\begin{eqnarray}
R_Q(D)&=&\min_{\theta}\left(\frac{\theta^2}{2J}+\max_{s\le
0}\left\{s\left(D-\frac{1}{2}\right)-\right.\right.\nonumber\\
& &\left.\left.\sum_xP(x)\ln\left[2\cosh\left(B+\theta-\frac{sx}{2}\right)\right]\right\}\right)
+\phi(B,J),
\end{eqnarray}
which requires merely optimization over two parameters. In fact, the
maximization over $s$, for a given $\theta$, can be carried out in closed form,
as it boils down to the solution of a quadratic equation. Specifically,
for a symmetric source ($P(-1)=P(+1)=1/2$), 
the optimum value of $s$ is given by
\begin{equation}
s^*=\ln\left[\sqrt{(1-2D)^2c^2+4D(1-D)}-(1-2D)c\right]-\ln[2(1-D)],
\end{equation}
where
\begin{equation}
c=\cosh(2B+2\theta).
\end{equation}
The details of the derivation of this expression are omitted as they
are straightforward.

As the Curie--Weiss model is well known to exhibit phase transitions (see,
e.g., \cite{Honerkamp02},\cite{MM09}), it is expected that $R_Q(D)$,
under this model, would consist of phase transitions as well. At the very
least, the last term $\phi(B,J)$ is definitely subjected to phase transitions
in $B$ (the magnetic field) and $J$ (the coupling parameter).
The first term, that contains the minimization 
over $\theta$, is somewhat more tricky
to analyze in closed form. In essence,
considering $s^*\equiv s^*(\theta)$ as a function of $\theta$,
substituting it back into the expression of $R_Q(D)$, and finally, 
differentiating  w.r.t.\ $\theta$ and equating to zero (in order to minimize),
then it turns out that 
the (internal) derivative of $s^*(\theta)$
w.r.t.\ $\theta$ is multiplied
by a vanishing expression (by the very definition of $s^*$ as a
solution to the aforementioned quadratic equation). The final result of
this manipulation is that the minimizing $\theta$ should be a solution 
to the equation
\begin{equation}
\theta=J\sum_xP(x)\tanh\left(B+\theta-\frac{s^*(\theta)x}{2}\right). 
\end{equation}
This is a certain (rather complicated) variant
of the well--known magnetization equation in the mean field model,
$\theta=J\tanh(B+\theta)$, which is well known 
to exhibit a first order phase transition
in $B$ whenever $J > J_c=1$. It is therefore reasonable to expect
that the former equation in $\theta$, which is more general, will
also have phase transitions, at least in some cases.

\section{Summary and Conclusion}
 
In this paper, we have drawn a conceptually simple analogy between
lossy compression of memoryless sources and statistical mechanics
of a system of non--interacting particles. Beyond the belief that this
analogy may be interesting on its own right, we have demonstrated its
usefulness in several levels. In particular, in the last section,
we have observed that the analogy between the information--theoretic
model and the physical model is not merely on the pure conceptual
level, but moreover, analysis tools from statistical mechanics can be
harnessed for deriving information--theoretic functions. Moreover, physical
insights concerning phase transitions, in systems with strong interactions,
can be `imported' for the understanding possible irregularities in these  
functions, in this case, non--smooth dependence on $B$ and $J$.\\

\end{document}